\documentclass[letterpaper]{article}

\usepackage[T1]{fontenc}

\usepackage{geometry}
\usepackage{setspace}
\usepackage[style = chem-acs]{biblatex}
\usepackage{graphicx}
\usepackage{float}
\newfloat{scheme}{htbp}{los}
\floatname{scheme}{Scheme}
\floatname{chart}{Chart}
\newfloat{graph}{htbp}{loh}

\usepackage{chemformula} 

\usepackage{authblk}
\author[1]{Caleb H. DeWitt}
\author[1]{Nnamdi C. Okey}
\author[1]{Griffin W. Hancock}
\author[1]{Aditi Bhattacherjee*}
\affil[1]{Department of Chemistry, University of Iowa, Iowa City, Iowa 52242, United States}

\title{Active Control of Extreme Ultraviolet Photon Flux and Resolution with a Plane Ruled Reflection Grating Spectrometer that Measures Diverging High-Harmonics}

\date{*Email: aditi-bhattacherjee@uiowa.edu}

\begin{document}

\maketitle

\begin{abstract}

High harmonic generation is a versatile tabletop source for the investigation of ultrafast dynamics with atomic fidelity and few-femtosecond time resolution. However, intensity modulations in the high harmonic spectra due to an alternating peak-and-valley structure in femtosecond extreme ultraviolet (XUV) sources can lead to significant distortions in measured XUV absorption spectra.
We report a broadband extreme ultraviolet (XUV) spectrometer with a plane ruled reflection grating that can be spectrally adjusted on the fly using a motorized iris aperture in the entrance slit to control the divergence of the XUV beam prior to its dispersive detection. The effective divergence is tuned between 0.03 to 1.9 milli-radian with concomitant variation in the detected XUV photon counts of 1.0 $\times$ 10$^6$ photons/s/eV to 1.5 $\times$ 10$^7$ near 40 eV.
Metallic samples of Fe (10 nm) and Ti (10 nm) deposited on Si$_3$N$_4$ membranes (100 nm) are measured under these conditions, with a spectral coverage of > 30 eV and a resolution of up to 500 meV. A volcano plot between signal-to-noise ratio (SNR) and spectral resolution determines the trade-off point in measuring elemental near-edge absorption spectra of thin film samples.
Consistent with ray tracing simulations, we show that this design parameter offers on-demand high resolution or high SNR configurations for the rapid acquisition of XUV absorption spectra of thin film solid-state samples, especially strongly correlated materials.

\end{abstract}

\newpage
\section{Introduction}

\begin{figure}
    \centering
    \includegraphics[width=1\linewidth]{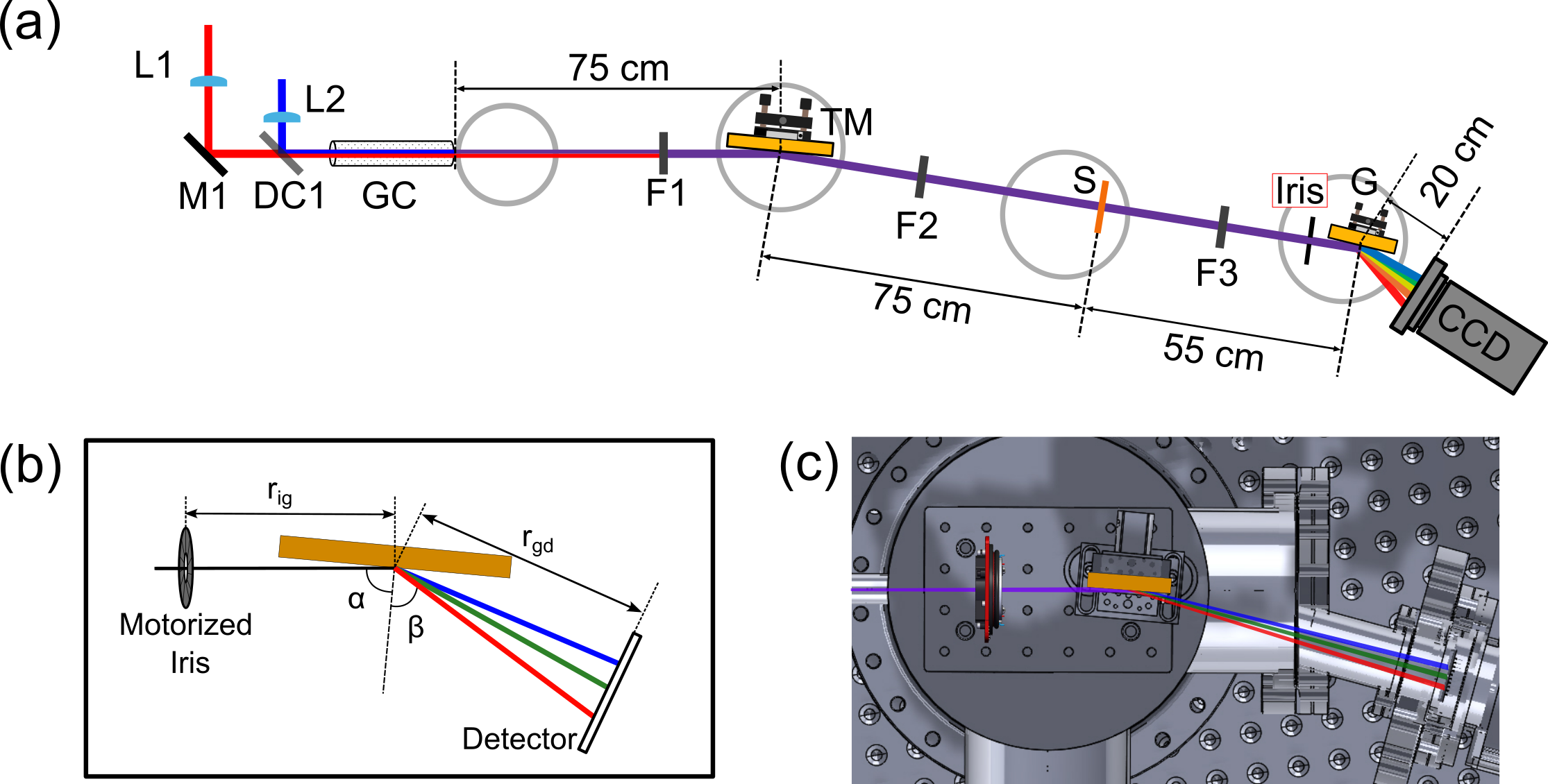}
    \caption{(a) Schematic of the XUV beamline and (b) detailed drawing of the spectrometer. Here L1: 75 cm focal length lens, L2: 50 cm focal length lens, GC: gas cell, F1,F2,F3: 200 nm aluminum foil filters, TM: toroidal mirror with 75 cm object and image plane distances, S: sample, G: 600 lines/mm plane ruled reflection grating, r$_{ig}$ is the iris to grating distance = 10 cm, r$_{gd}$ is the grating to detector distance = 20 cm. (c) A CAD drawing of the spectrometer chamber is shown for comparison.}
    \label{fig:Beamline}
\end{figure}

High-harmonic generation (HHG) provides a tabletop, tunable, extreme-ultraviolet (XUV) and soft X-ray source with femtosecond-to-attosecond temporal resolution. With these broadband and short-pulse capabilities, tabletop HHG sources have provided valuable insight into coupled electron and nuclear dynamics in atoms,\cite{loh2013, young2018} molecules,\cite{bhattacherjee2018, carlson2025, vura2025} and solids.\cite{geneaux2019, biswas2022}
Generally, high harmonic generation uses strong-field ionization of a rare gas medium contained in a gas cell with a near-infrared driving laser that is focused (Figure \ref{fig:Beamline}a) at an intensity of > 10$^{13}$ W/cm$^2$. The resultant XUV photons constitute several narrow bandwidth peaks that are higher-order harmonics of the driving laser frequency. In the case of a typical Ti:Sapphire femtosecond amplifier (with an 800 nm central wavelength or 1.55 eV), HHG manifests itself as narrow bandwidth peaks separated by $\approx$3.10 eV. Only odd order harmonics are produced as the induced polarization vanishes for even orders of the electric field in centrosymmetric generating media such as atoms.

Alternatively, when the inversion symmetry of the 800 nm field ($\omega$) is broken by a spatio-temporally coinciding second harmonic (2$\omega$, 400 nm) field, even-numbered harmonics are generated.\cite{ehlotzky2001,pfeifer2006}
Regardless of the generation scheme (single color, 800 nm, or two-color, 800 nm + 400 nm) and driving laser wavelength (800 nm of Ti:Sapp or 1030 nm of Yb lasers), the HHG spectrum comprises several high photon-flux peaks separated by low photon-flux valleys that can span several tens of electron volts from the high ponderomotive energies that a free electron acquires in an intense laser field.\cite{ding2014, lorek2014}
The alternating low- and high-flux spectrum creates intrinsic modulations in the measured static absorption spectra of gas and solid samples due to many factors.\cite{attar2014, lin2016, yang2018}

\begin{figure}
    \centering
    \includegraphics[width=1.0\linewidth]{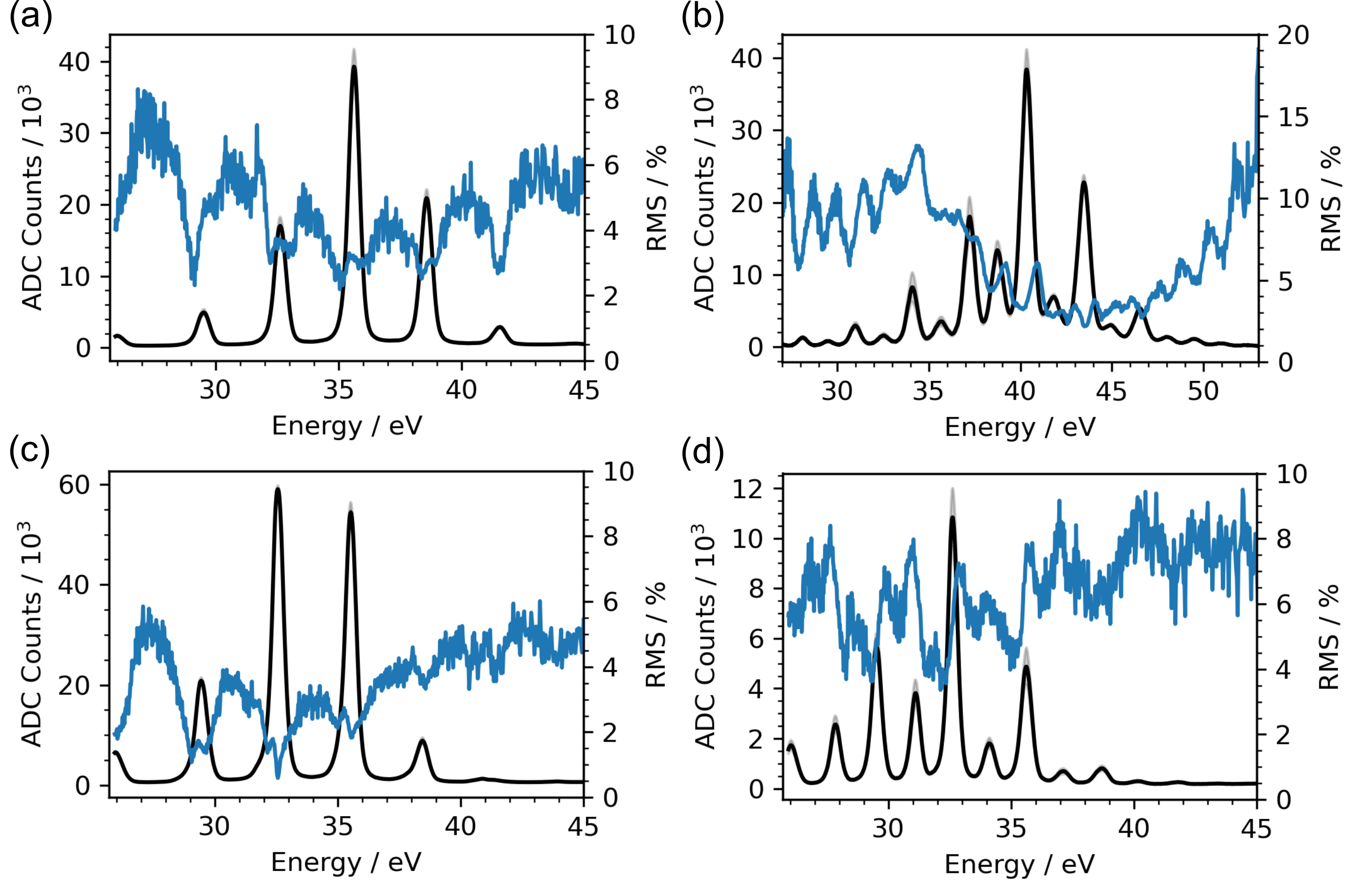}
    \caption{High harmonic spectra (black trace) generated in argon gas and accompanying standard deviation over 1000 pulses (gray shaded region). The root mean squared (RMS) fluctuation is plotted as a percentage and shows periodic modulations as described in the text. Panel (a) is generated at 30 torr of argon and collected with a 30 ms exposure. (b) 40 torr argon at 100 ms exposure with both even and odd ordered harmonics. (c) 22 torr argon at 150 ms exposure. (d) 21 torr and 40 ms exposure with even and odd ordered harmonics.}
    \label{fig:argonstability}
\end{figure}

First, a lower flux in the valleys compared to the peaks adversely affects the signal to noise ratio (SNR), or, in the best-case scenario, produces non-uniform fluctuations throughout the spectrum (Figure \ref{fig:argonstability}).\cite{dewitt2026}
Second, measurement factors such as detector nonlinearity, stray light scatter, and CCD (charge-coupled device) readout noise affect the regions of the absorption spectrum in a proportional manner to the total photon counts, again affecting the high- and low-flux regions to a different extent.\cite{zadnik1998, zong2006, zonios2010} Consequently, the high harmonic structure often manifests itself as modulations in the absorption spectrum following the pattern of the harmonic peaks. As these contributions are not caused by random (gaussian) noise, they cannot be mitigated by simply increasing the averaging of laser pulses in data acquisition schemes.
These effects are further exacerbated when studying thin films, as the total flux is severely diminished by the substrate on which the samples are deposited (typically thin silicon nitride < 200 nm, Si$_3$N$_4$ membranes).

In a typical HHG experiment, the divergent XUV source is refocused in a sample by a grazing incidence toroidal mirror (Figure \ref{fig:Beamline}a). After interaction with the sample, the propagating XUV beam is diverging and detected by an energy dispersive spectrometer, generally a grazing-incidence reflective-type diffraction grating and a CCD or MCP (microchannel plate) imaging screen.
To mitigate a lowering of the spectral resolution due to angular divergence of the XUV beam, typical XUV spectrometers  will at least partially refocus the beam with a second toroidal mirror,\cite{loh2008, li2025} or by using a concave spherical grating capable of (again, partially) refocusing in the sagittal plane while dispersing the harmonics in the tangential plane.\cite{wang2013, wanie2024, li2024}
Strong focusing is not possible due to limitations on the curvature of a substrate while maintaining the requisite surface flatness for XUV wavelengths.
Additionally, every  gold-coated optic used at grazing incidence cuts down the photon flux by nearly $\approx$ 10-20\% in the XUV between 30 and 60 eV.\cite{dewitt2026}

We present an XUV spectrometer that uses a divergent incoming XUV beam and implements a motorized iris to provide on the fly control of the spectral resolution and overall flux of detection (Figure \ref{fig:Beamline}b-c). By tuning the spectral resolution, valleys in the harmonic spectrum are filled by overlapping harmonics at the cost of resolution, enabling the fast acquisition of high signal-to-noise absorption spectra in metallic thin film samples without resorting to any post-correction algorithms.\cite{facciala2021, geneaux2021}
We show that moderate resolution ($\sim$ 500 meV) and high-flux configurations provide high signal-to-noise spectra while preserving the overall shape of the core-to-valence electronic absorption spectrum.
We collect both the giant resonance, correlated XUV absorption spectrum of a Ti metal film near 45 eV as well as the Fermi-level absorption in a Fe thin film near 55 eV. A volcano plot at the iron M-edge of the SNR versus effective XUV half-angle divergence is used to demonstrate the trade-off point between these two parameters (SNR and divergence) with grating spectrometers that measure diverging high harmonics. The method is expected to boost the acquisition of XUV absorption spectra of strongly correlated materials with a high SNR.

\section{Experimental}

The output of a Legend Duo Elite, Coherent, Inc., (>13 W, <35 fs, 1 kHz, 800 nm central wavelength) regenerative amplifier seeded by a Vitara-S oscillator (800 nm central wavelength, 600 mW, >70 nm bandwidth, <20 fs at 80 MHz) is used to drive high harmonic generation. The dual amplifier (regenerative first stage plus single pass second stage) laser system is capable of producing >13 mJ, sub-35 fs pulses at a 1 kHz repetition rate; however, for XUV generation discussed here the timing of the green pump (Revolution 80) in the single pass amplifier is intentionally mismatched. This scheme allows for regenerative amplification in the first stage without amplification in the second stage while maintaining the thermal lens effect for good mode quality, effectively reducing the amplified output power by half (> 8 W) before final compression.

The amplified output is split by an 80:20 beam splitter, and the two unequally split beams are compressed independently with the weaker beam (1.6 W) routed through the internal compressor inside the amplifier and the stronger beam (6.5 W) propagated outside the amplifier into an external compressor.
The stronger split-off beam that is compressed externally is further routed by a 90:10 beam splitter where the stronger arm (2.5 W) is focused into a semi-infinite gas cell using a 75 cm FL fused silica lens to drive HHG. The weaker component (0.5 W) is down-collimated and passed through a 500 $\mu$m $\beta$-BBO crystal cut at 29$^\circ$ for second harmonic generation ($\sim$ 45 $\mu$J). The polarization of the second-harmonic beam is rotated by a $\lambda/2$-wave-plate and focused by a 50 cm FL lens into the semi-infinite gas cell. This two-color, 800 nm + 400 nm co-propagating pulse scheme enables the generation of even and odd harmonics. A micrometer-stage-mounted retroreflector provides fine control to ensure temporal overlap of the two pulses.

The XUV beamline comprises four separate stainless steel vacuum chambers connected by manual gate valves and 200 nm Al foils mounted on manual push-pull feedthroughs for filtering of the residual near-IR fundamental used to drive HHG.
The setup is described in detail elsewhere,\cite{dewitt2026} while noting here that there are only two XUV optics utilized - a one-to-one imaging toroidal mirror (75 cm focal length) and a plane ruled reflection grating (600 lines/mm, Figure \ref{fig:Beamline}a).
XUV static absorption spectra are collected for 10 nm films of iron and titanium coated on 100 nm silicon nitride membranes (Silson Ltd) by averaging high harmonic spectra transmitted through the sample film and referencing the average of 64, 128, or 256 spectra of a blank 100 nm silicon nitride window. Exposure times (between 200 ms and 2 s) are chosen until the measured spectrum through the blank silicon nitride reads between 35\,000 and 45\,000 counts on the detector. Spectra are background corrected in the Andor Solis software suite before each measurement, which accounts for stray light and thermal noise in the detector (cooled to -25 $^\circ$C). The absorbance of the sample is calculated from the averaged background-corrected spectra as $A = -\log_{10}(I_{sample}/I_{ref})$.

Ray-tracing simulations are performed using the SHADOW3 code.\cite{sanchez2011} The source is constructed as a 65 $\mu$m point source, calculated from the expected beam waist of our focused NIR beam.
The simulated spectra comprise 1 eV FWHM harmonics that are separated by 1.55 eV in the range of 35 to 72 eV.
The divergence is set with a half-angle divergence of 813 $\mu$rad, as determined from the vertical footprint of the harmonics on the CCD in the experimental setup. A toroidal mirror is specified with a source and image plane distance of 75 cm, similar to the experiment, and the radii of curvature are set to match the specifications of the actual optic. A screen with a circular aperture is simulated 45 cm from the previous image plane. Finally, a 600 l/mm grating is simulated 10 cm from the aperture with an incidence angle of 85.1$^\circ$ and an image plane of 20 cm to closely match the location of the grating and the CCD in our experimental setup.

\section{Results and Discussion}

\subsection{Ray Tracing Simulations}

\begin{figure}
    \centering
    \includegraphics[width=1.0\linewidth]{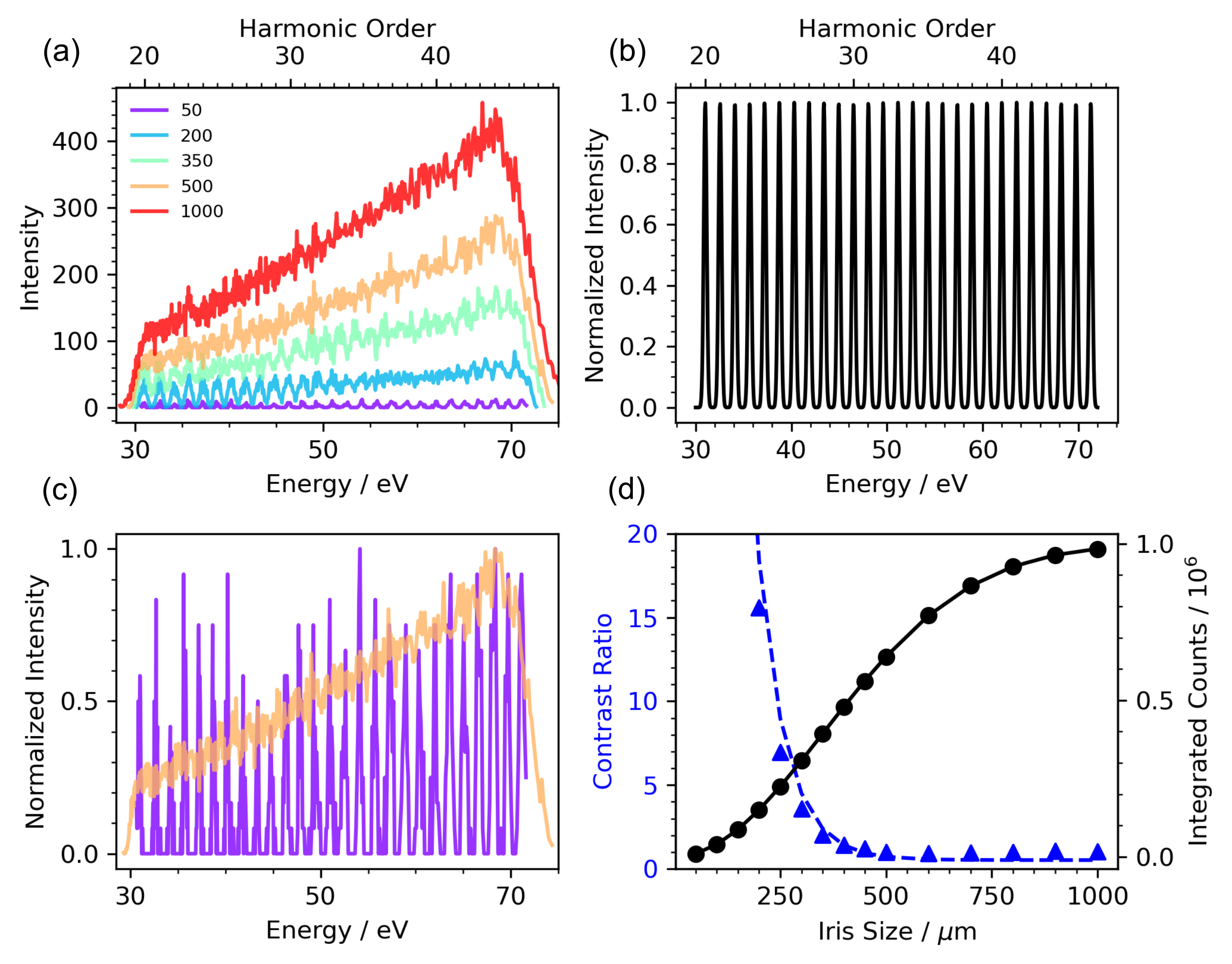}
    \caption{(a) Simulated harmonic spectra at varying iris sizes (50 to 1000 $\mu$m) using SHADOW3 ray tracing. (b) The input spectrum used as the photon source. (c) Two example spectra (50 $\mu$m and 500 $\mu$m) in (a) are normalized for direct comparison of peak shapes. (d) The integrated flux is plotted against the iris size (black circles) and then fit (black trace) to obtain a gaussian width of $\sigma$ = 175 $\mu$m. The contrast ratio for harmonic 21 (32.55 eV) is shown as the blue triangles and fit to an exponential (blue dashed trace) to guide the eye.} 
    \label{fig:RayTracing}
\end{figure}

Figure \ref{fig:RayTracing}a shows the results of ray-tracing simulations of high harmonic spectra with varying aperture diameter between 50 $\mu$m and 1 mm in the entrance slit. A rendering of the design apparatus used for ray tracing is shown in Figure S1. At the source, a constant flux of high harmonics (totaling 10$^6$ rays, Figure \ref{fig:RayTracing}b) is specified across all harmonic orders between 19 to 45, which corresponds to photon energies of 30 eV and 70 eV, respectively. 
This energy range is chosen for a direct comparison with our tabletop HHG source, which is equipped with Al filters (L$_{2,3}$ edge = 72 eV) on push-pull feedthroughs to remove residual infrared.
The source spectrum and bandwidth per simulated harmonic are shown in Figure S2.
The spatial distribution of the XUV in the detection plane along the dispersion axis is mapped along the horizontal pixel axis of the CCD (Figure \ref{fig:RayTracing}a, c) which is converted to energy using the grating equation in first-order diffraction. In real space, it spans 25 mm in length, which is the approximate length of the CCD chip (2048 pixels $\times$ 13 $\mu$m pixel size, or 26.624 mm) deployed in the experimental apparatus.

Detected XUV photons in the simulations are seen to bunch at higher energies because of a lower resolving power for a fixed groove density of the grating, as also commonly seen in experiments (\textit{vide infra}).
The upper and lower limits of the sampled aperture size are normalized and plotted directly below in Figure \ref{fig:RayTracing}c. The noise in the detected photon counts in our simulation is due to Monte Carlo sampling of rays (10$^6$ in a single run), and can be mitigated by using more advanced simulation tools such as loops.\cite{rebuffi2016}
Notwithstanding the noise in the simulations, the flux drops off dramatically with the size of the iris as $\approx 1/r^2$, and the harmonic peak widths become progressively narrower to provide a high spectral contrast between the peaks and the valleys. 

The integrated photon counts at the detection plane are shown in Figure \ref{fig:RayTracing}d. In the open aperture limit, it adds to the full 10$^6$ rays that are simulated. Between 250 and 500 $\mu$m, a linear change in intensity is registered, and the counts begin to plateau above 600 $\mu$m entrance aperture size.
It accurately fits an inverse gaussian distribution, $I(r) = A(1-e^{-r^2/2\sigma^2})$ with $\sigma$=175 $\mu$m, where the functional form noted represents the integral of a gaussian in polar coordinates.
High harmonic peaks are indistinguishable from the valleys in this domain (> 600 $\mu$m), resulting in a peak-to-valley contrast ratio of one.
The loss in periodic structure at the detection plane is attributed to the increasing overlap between neighboring harmonic orders as a result of the intrinsic divergence of the XUV rays.
The spectral contrast improves rapidly, up to $>$15 at 150 $\mu$m aperture size; exceeding $>>$100 at smaller aperture size as the flux in the valleys tends to zero. Below, we demonstrate the interplay between high harmonic flux and contrast on our tabletop source, its agreement and variation from ray tracing simulations at the diffraction limit, before finally rationalizing its impact on experimental SNR measurements at the pre-edge.

\subsection{Experimental High Harmonic Images - XUV Beam Divergence}

\begin{figure}
    \centering
    \includegraphics[width=1.0\linewidth]{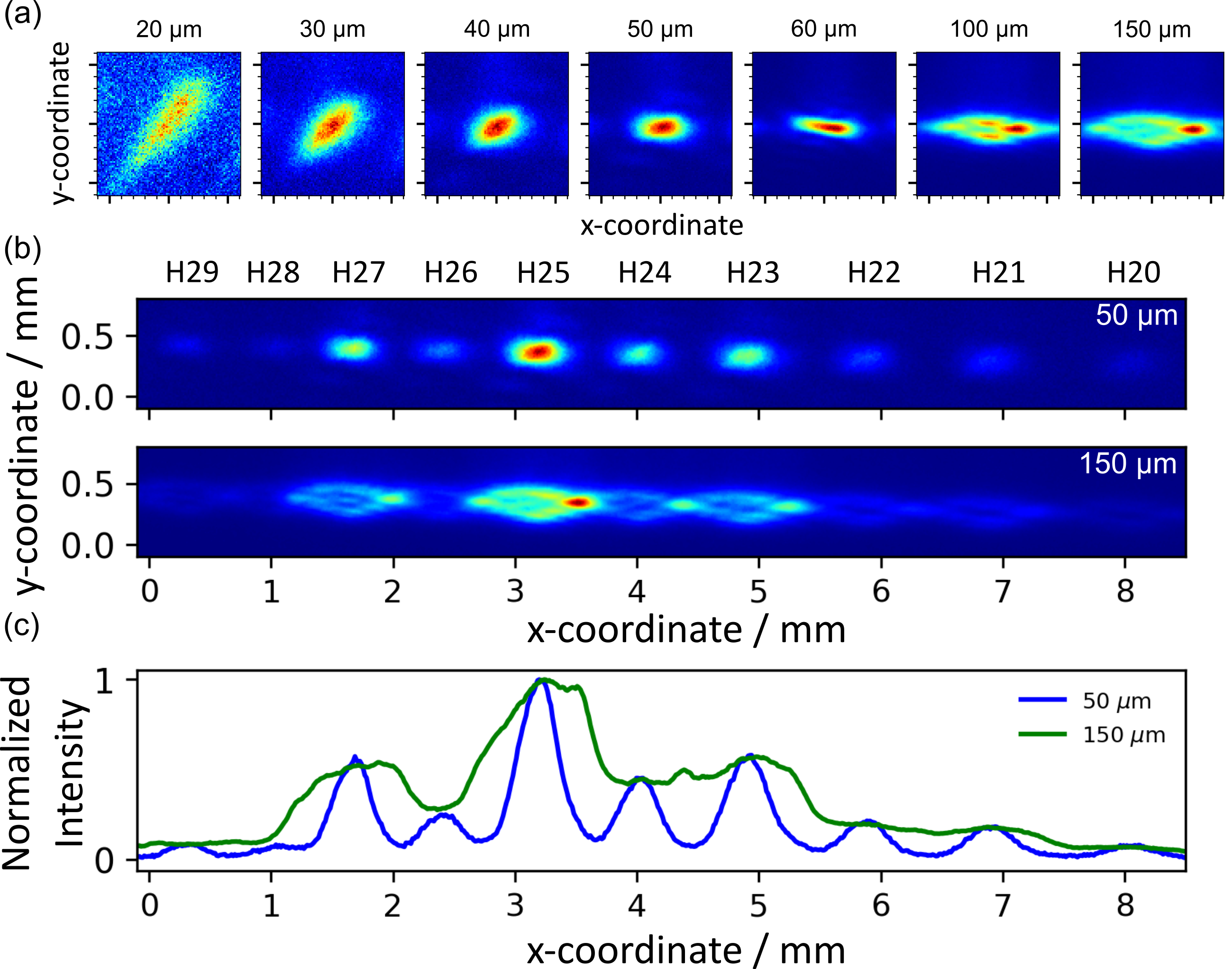}
    \caption{(a) Images of an isolated harmonic (H25, 38.8 eV) on the CCD shown for various iris sizes (20 to 150 $\mu$m). Here, the large ticks represent a spacing of 0.5 mm such that the individual frames are 1.2 $\times$ 1.2 mm in spatial extent (x $\times$ y). (b) Images of the full harmonic spectra in argon at two representative iris sizes (50 $\mu$m and 150 $\mu$m). (c) Integrated spectra of (b) obtained by full vertical binning and normalized.}
    \label{fig:Divergence}
\end{figure}

In our implemented XUV beamline shown in Figure \ref{fig:Beamline}a, focusing of the beams occurs twice - that of the driving infrared (by lens L1) in the gas cell and later downstream of the XUV by the toroidal mirror at the position of the sample. A knife-edge measurement of the XUV spot size at the focus (where the sample is placed) returns 140 $\pm$ 20 $\mu$m for the XUV beam waist. The transmitted beam is divergent from this point and dispersed by the plane ruled reflection grating on the CCD. An image of a representative (25th) harmonic order as a function of aperture size is shown in Figure \ref{fig:Divergence}. We note that the entrance slit aperture is not perfectly circular (Figure S3) and creates an asymmetric shape for some values of the aperture size. With the iris fully open and not constricting the diverging XUV beam, the beam divergence on the vertical axis is measured to be 2 mm, consistent with both the ray tracing simulations and the calculated numerical aperture of the focusing lens L1.

Controlling the iris down to 100 $\mu$m in diameter allows one to tune the projected XUV beam divergence. Below 100 $\mu$m, the diffraction effects become significant, \textit{vide infra}. Figure S4 shows a comparison of high harmonic spectra at the threshold of diffraction and its adverse effect on the high harmonic peak widths as well as spectral contrast. Theoretically, the highest resolution derives from the single ray that passes through the center of the iris. However, in practice, the diffraction limit beats the theoretical upper bound on resolution. Figure \ref{fig:Divergence}b shows a section of the CCD chip (650 by 58 pixels) for high harmonics generated in argon between 30 and 45 eV. Two representative entrance aperture diameter values of 50 $\mu$m and 150$\mu$m are highlighted. The corresponding spectra measured upon full vertical binning, and normalized to one, are shown directly below for comparison. The divergence is comparable to the values reported for other tabletop HHG sources driven by ultrashort near-infrared pulses near 800 nm.\cite{wong2010, von2016, fu2020} This level of agreement is expected, as the divergence is primarily determined by the focusing conditions and the distance from the point source.\cite{fu2020} Maintaining optimum phase matching conditions along the propagation axis shows a gaussian-like profile of the high harmonics.\cite{takahashi2004, fu2020} The gas medium and the high harmonic order also play a significant role.\cite{wong2010}

\subsection{Experimental High Harmonic Spectra - Scaling and Contrast}

\begin{figure}
    \centering
    \includegraphics[width=1.0\linewidth]{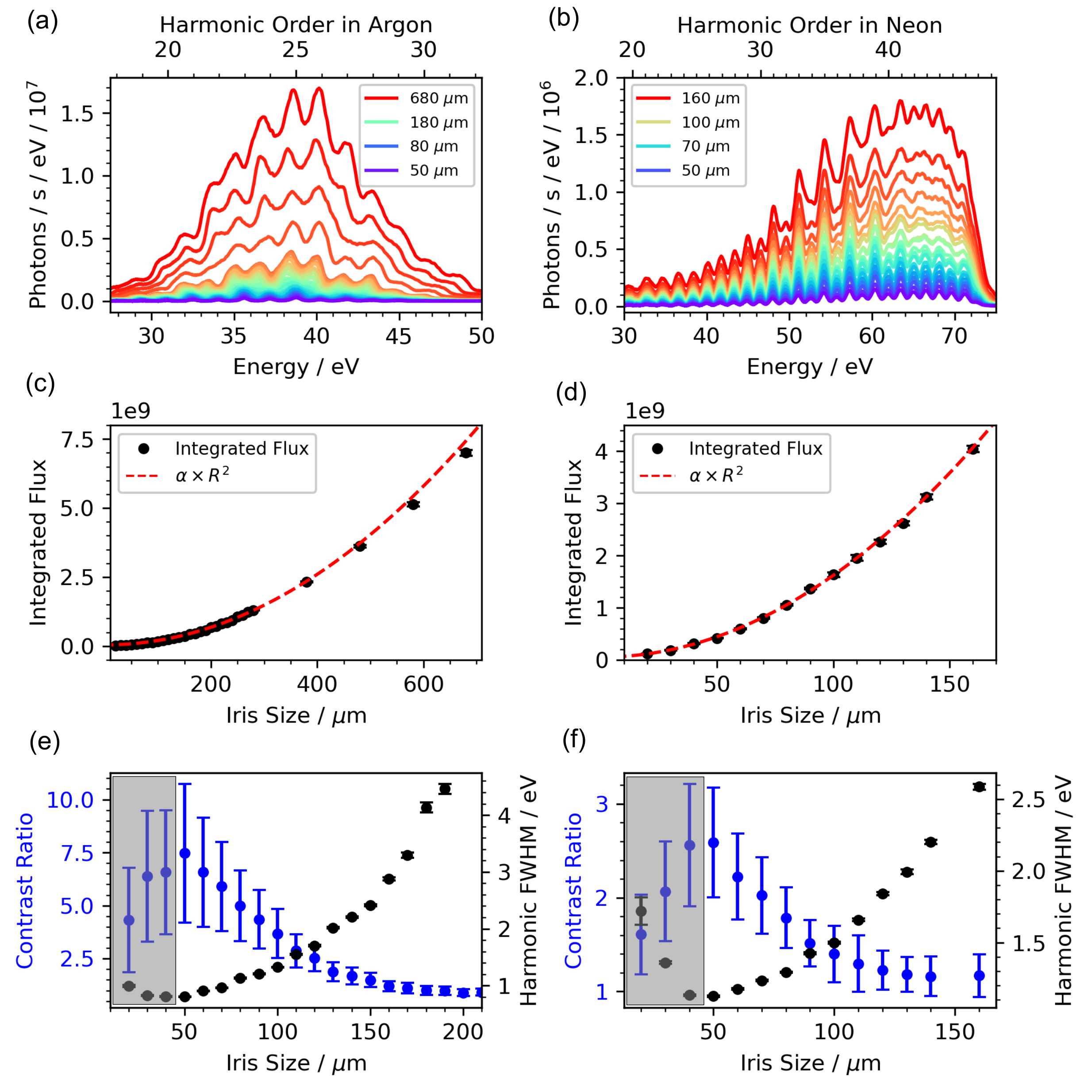}
    \caption{Harmonic spectra generated in argon (a,c,e) and neon (b,d,f) with varying iris aperture diameter in the entrance slit.
    The integrated flux is plotted (c and d) and fit to a scaled $\alpha \times R^2$ curve where the scaling factor, $\alpha$, is 60\,897 photons/s/eV/$\mu$m$^2$ for the argon harmonics and 622\,144 photons/s/eV/$\mu$m$^2$ for the neon harmonics. A central prominent harmonics (H25 at 37.3 eV for argon and H39 at 60.3 eV for neon) is fit to a gaussian, and the full-width at half-maximum (FWHM) of the fit is plotted against the iris size (e and f, black circles) along with the contrast ratio (blue circles, error bars denote one standard error).} 
    \label{fig:HarmonicSpectra}
\end{figure}

High harmonic spectra generated by a two-color 800 nm + 400 nm laser field in argon (25 torr) and neon (100 torr) are shown in Figure \ref{fig:HarmonicSpectra}a-b at varying sizes of the iris aperture in the entrance slit. At 24 torr of Argon with $\sim$ 1.0 mJ of a 35 fs, 800 nm driving pulse overlapped with 50 $\mu$J of the 400 nm second harmonic, harmonics spanning 28 to 45 eV emerge.
Harmonic spectra generated in 95 torr neon with the same driving conditions produce energies ranging from 31 eV to the 72 eV energy cutoff due to the aluminum filters (Figure \ref{fig:HarmonicSpectra}c and d).
For a more direct comparison of the high harmonic spectral profile and full width at half maximum (FWHM) at varying half-angle divergence values, the normalized data are shown in Figure S4. The flux shown in Figure \ref{fig:HarmonicSpectra}a-b is measured at the detector and does not account for losses due to transmission through filters or the reflectivity of the gold-coated toroidal mirror and grating. The detected photon count is expected to be about 10$\%$ of what is generated at the source.\cite{bhattacherjee2016, schnorr2019, ash2023}
The XUV photon flux is comparable to other implementations using Ti:Sapphire lasers,\cite{loh2008, ding2014} and brighter flux exceeding 10$^{12}$ photons s$^{-1}$ is reported at higher repetition rates.\cite{hadrich2014, wang2015}

Closing the iris at the entrance slit effectively reduces the angular divergence of the incident XUV, enhancing the resolution and spectral contrast.
For example, when the iris is mostly open (680 $\mu$m, Figure \ref{fig:HarmonicSpectra}a), the spectrum appears as a broader continuum of overlapping harmonics with discernible periodicity in the structure. As the iris aperture size is restricted, the underlying structure of sharp harmonics emerges with a concomitant drop in the extreme ultraviolet photon flux. In fact, the presence of even harmonics with the use of the symmetry-breaking field can only be perceived below an entrance slit diameter of 150 $\mu$m.
Similar observations are made with neon as the generation medium (Figures \ref{fig:HarmonicSpectra}b and S2). It is important to note that the grating equation naturally dictates a smaller angular separation at higher energies, making the harmonics in neon appear closer together and arising from a broad pedestal, an effect that is also captured in the ray tracing simulations discussed above. This limitation prevents a reliable extraction of harmonic widths beyond 200 $\mu$m to obtain the contrast ratio in neon, but does not affect the analysis shown in Figure \ref{fig:HarmonicSpectra}f.

The total counts detected for each spectrum are plotted against the iris size in Figure \ref{fig:HarmonicSpectra}c-d. At smaller iris sizes (< 400 $\mu$m), the trend follows closely with a quadratic ($r^2$) curve. This trend is expected as the area of the iris aperture increases proportionally to $r^2$. However, for a gaussian beam profile like this one, this trend only holds in the region where the intensity of the gaussian remains largely invariant, and thus the intensity increases proportional to the size of the iris at small apertures. However, at larger aperture sizes, the true values deviate marginally from the $r^2$ curve as the aperture reaches the edges of the gaussian profile.
As nominally employed in a prototypical knife edge measurement of the beam waist, the increase in photon flux mimics the integration of the gaussian intensity profile.
However, in the case of a circular iris, the integrand is, in essence, over the radius of a circular gaussian in polar coordinates $I(r,\theta) = Ae^{-r^2/2\sigma^2}$, where $A$ is the amplitude and $\sigma$ is the standard deviation. As noted earlier, this integral takes the form of an inverse gaussian as $\int Ae^{-r^2/2\sigma^2}rdr = A(1-e^{-r^2/2\sigma^2})$.

A central and prominent harmonic from the spectrum is chosen and fitted to a gaussian function to determine the variation in peak width with iris size (Figure \ref{fig:HarmonicSpectra}c). Data are only fit up to an iris size of 250 $\mu$m, after which the overlap of neighboring harmonics provides poor fits, risking an inaccurate representation of the true harmonic width. As expected, the harmonic width decreases with decreasing iris size before reaching a minimum near 50 microns. Below this value, the harmonic widths begin to increase due to diffraction, which is shown in the gray box region in Figure \ref{fig:HarmonicSpectra}e-f. In the nearby range (40 $\mu$m - 60 $\mu$m), the harmonic width is relatively unchanged as the maximum resolution of the grating is reached (i.e., the true energy width of the harmonic peaks). Below this point, the peak widths begin to widen, which is attributed to increased diffraction from the smaller aperture size. This behavior highlights a key limitation of the method, \textit{i.e,} diffraction due to a small size of the aperture sets a lower limit to the usable iris size and the maximum achievable resolution. The diffraction-limited resolution is not captured in the ray tracing simulations and is further shown in the bottom panels of Figure S4.

\subsection{Static XUV Absorption Spectra}

\begin{figure}
    \centering
    \includegraphics[width=1.0\linewidth]{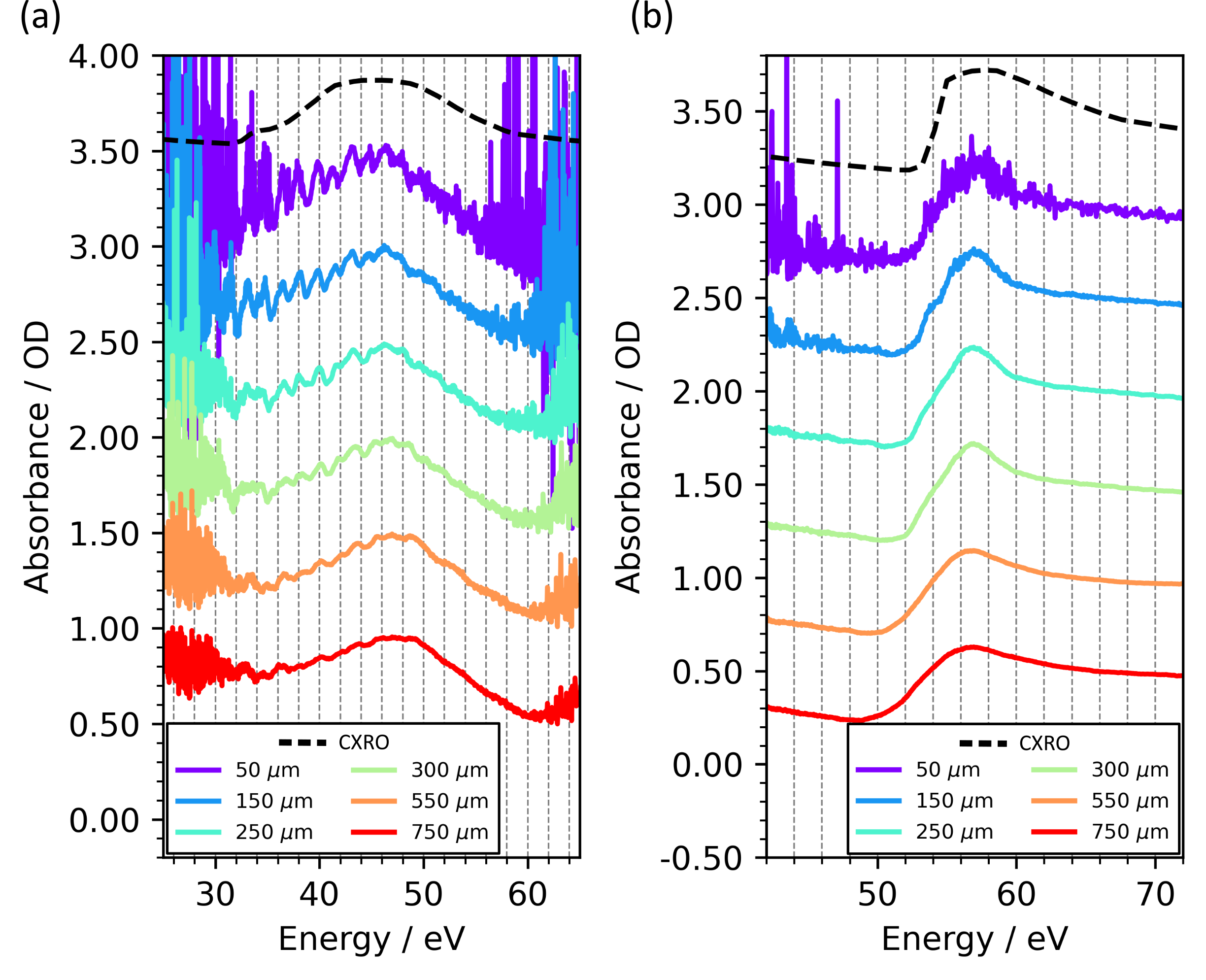}
    \caption{Absorption spectra of a 10 nm film of (a) iron and (b) titanium measured at varying iris sizes. The spectra are vertically offset for clarity. The optical density obtained from CXRO of 15 nm of iron and 7 nm of titanium are shown as the dashed black trace at the top.}
    \label{fig:StaticAbsorptionSpectra}
\end{figure}

The high harmonics generated in argon and neon are used to measure the M$_{2,3}$ absorption edges in thin metallic films of titanium and iron deposited on free-standing Si$_3$N$_4$ membranes using electron beam deposition. Figures \ref{fig:StaticAbsorptionSpectra}a and \ref{fig:StaticAbsorptionSpectra}b plot the near-edge absorption spectra with varying iris size to analyze the effects of angular divergence on the shape and spectral resolution in a typical XANES measurement.
We note that the Fe M$_{2,3}$ edge registers a sharp near-edge resonance expected for a typical XANES spectrum. The near-edge feature is identified at the expected 52.7 eV corresponding to the energy of a Fe(3p)$\rightarrow$Fe(3d) transition.\cite{de2024} However, the Ti M$_{2,3}$ edge shows a giant resonance that manifests itself as a broad $\sim$30 eV feature centered around 45 eV.\cite{volkov2019_natphys,jansen2016} This feature is also attributed to Ti(3p)$\rightarrow$Ti(3d) transitions; however, it is significantly broadened due to a local field (screening) effect that arises from many-body interactions in a strongly correlated electronic system. 
Ti and Fe are well studied metals in the literature on XUV spectroscopy,\cite{jansen2016, volkov2019_natphys, de2024} including oxides,\cite{vura-weis2013, husek2017, carneiro2017, hruska2022} sulfides,\cite{nyrow2014} selenides,\cite{heinrich2023, huber2024} and heterojunctions.\cite{cushing2020}

The measured spectra (purple through red traces, Figures \ref{fig:StaticAbsorptionSpectra}a-b) are found to be in agreement with the spectrum reconstructed from the CXRO database (dashed black trace).
In the case of Ti, the smallest iris size shows the expected broad resonance peak shape; however, it is noisy due to severe modulations that align with the alternating structure of the harmonic spectrum (Figure S5).
The noise measured at the edges of the spectrum (< 35 eV and > 55 eV) is due to the limited bandwidth of the high harmonics in argon.
In addition, since the spectral resolution is inherently higher in argon harmonics compared to that in neon harmonics (as discussed earlier), the increased separation of individual harmonics contributes more strongly to these artifacts in Ti, associated with the alternating high- and low-flux peak structure. As the iris aperture widens (purple through red), the high harmonic modulations subside significantly with improved signal-to-noise ratio (SNR).

In the case of the iron M$_{2,3}$ edge (Figure \ref{fig:StaticAbsorptionSpectra}b), a distinct power law baseline from non-resonant ionization of core electrons can be identified between 42 eV to 48 eV, as also reflected in the CXRO data. We note that these spectra are referenced to 100 nm blank Si$_3$N$_4$ membranes such that the iron atoms are the major contributor to the power law baseline that runs from 0.3 units of optical density (at 42 eV) to 0.2 OD near 48 eV.
With gradual increase in the size of the iris aperture, the sharp onset and fall that characterize the iron edge at halfway marks of 54 eV and 58 eV, respectively, remain consistent up to an iris size of 250 $\mu$m, after which the near-edge feature begins to broaden indicating lower spectral resolution. From the peak widths measured in Figure \ref{fig:StaticAbsorptionSpectra}b, we estimate that the resolution has decreased from $\sim$600 meV to approximately 1.8 eV in the sampled range of iris sizes (80-250 $\mu$m) without affecting the location of the near-edge peak at 57.2 eV.

\subsection{Trade-off between Spectral Resolution and SNR at the Fe M$_{2,3}$ edge}

\begin{figure}[H]
    \centering
    \includegraphics[width=1.0\linewidth]{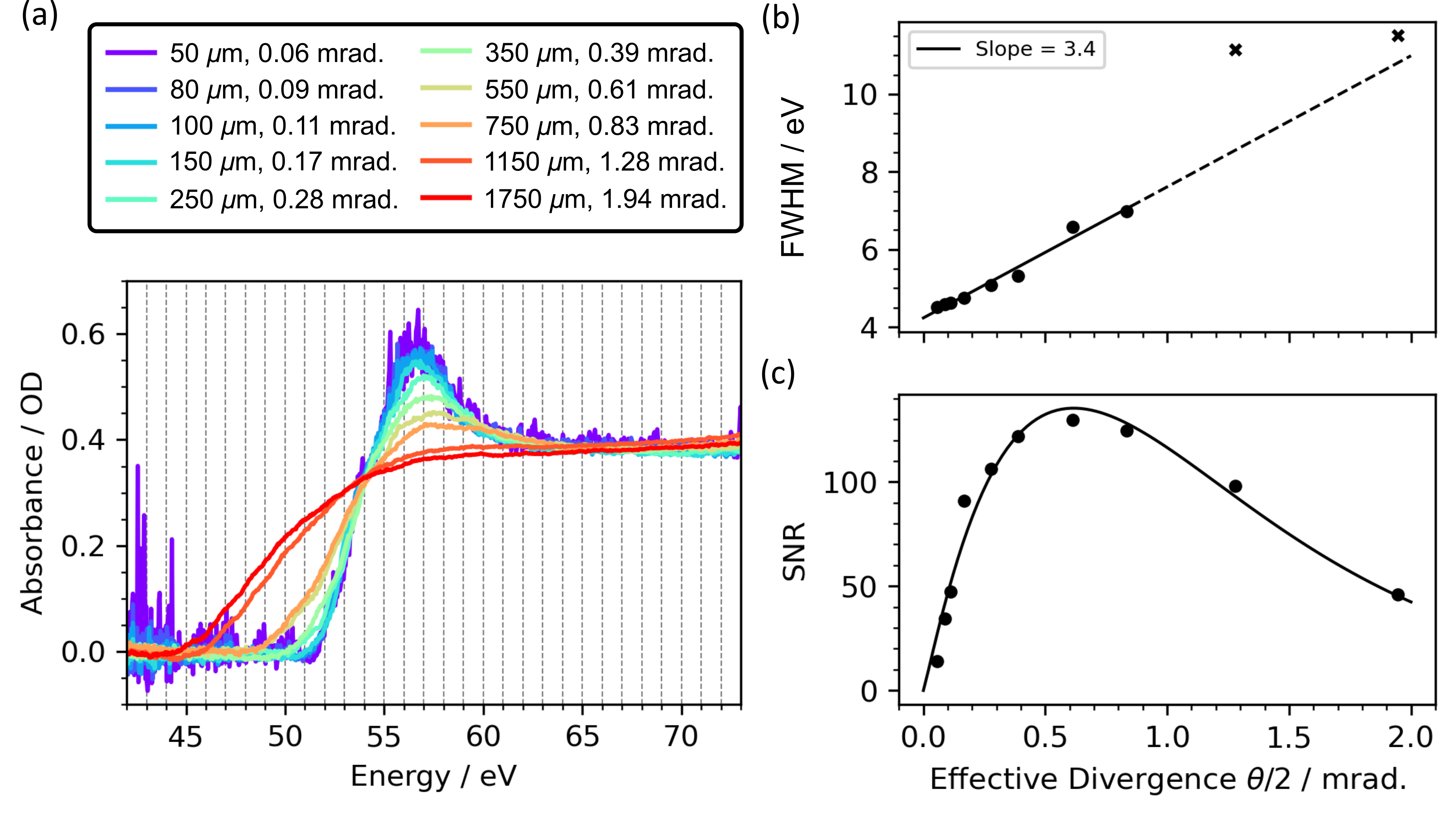}
    \caption{(a) XUV absorption spectrum at the iron M$_{2,3}$ edge of a 10 nm film at increasing iris sizes. The iris size is converted to an effective angular divergence ($\theta/2$) as indicated in the legend. The edge is fit to an error function combined with a gaussian and the determined FWHM is plotted against the effective divergence in (b) and fit to a line with a slope of 3.4 eV/mrad. (c) The signal-to-noise ratio (SNR) of each spectra is plotted against the effective angular divergence and fit to a function of the form $f(E) = A\times x^n\times e^{-x/\tau}$ to guide the eye.}
    \label{fig:SNR}
\end{figure}

The iron M$_{2,3}$ edge is a good model for demonstrating the behavior of the near-edge resonance peak width as a function of the entrance aperture size to identify the tradeoff between SNR and spectral resolution in tabletop XANES measurements.
Note that the acceptance cone angle of the iris aperture is greater than the intrinsic milli-radian divergence of the XUV source.
This allows us to track the near-edge M$_{2,3}$ resonance of iron in a systematic way. At narrow slit widths the resonance is comparable to the CXRO database (purple and blue traces, Figure \ref{fig:SNR}a), whereas in the extreme case only a rising edge is measured and the near-edge resonance is completely washed out (orange and red traces, Figure \ref{fig:SNR}a).
Here, we adopt the definitions outlined by St\"ohr for SNR, background and noise in experimental XANES / EELS (electron energy loss spectroscopy),\cite{stohr1992} as further exemplified in Figure S6.

The ionization edge in Figure \ref{fig:SNR} is modeled using an error function that is suitable to describe a gaussian density of states at the Fermi edge.\cite{paasch2010, jang2017} The error function utilizes a floating broadening parameter and a fixed inflection point. The broadening parameter is applied symmetrically with respect to the inflection point such that the center of gravity of the ionization energy remains unchanged (Figures S7-S12).
The full extent of the near-edge absorption (between 45-70 eV) is modeled using the sum of a gaussian with an error function.\cite{stohr1992}

More specifically, the error function utilized has the form $H\left[\frac 1 2 + \frac 1 2 \text{erf}\left(\frac{E-P_E}{\Gamma/c}\right)\right]$ where $H$ is the height, $P_E$ is the inflection point, $\Gamma$ is the FWHM of the error function, and $c=2\times \sqrt{\ln 2}=1.665$. The full functional form is, thus, $A\times \exp (-\frac{(E-P_G )^2}{\Gamma/c})+H\left[\frac 1 2 + \frac 1 2 \text{erf}\left(\frac{E-P_E}{\Gamma/c}\right)\right]$, where $A$ indicates the amplitude of the Gaussian and $P_G$ is its midpoint.
The FWHM versus entrance slit aperture returns a slope of 3.4 eV/mrad (3.8 meV/$\mu$m, or 3.8$\times$10$^{-3}$ eV/$\mu$m). The resolution of the spectrum depends on the effective angular divergence of the beam from the point of the iris (resolution $\propto \tan(\theta)$). In the effective angular divergence expected for the aperture sizes studied (up to 2 mrad), the small angle approximation holds such that a linear relationship adequately captures the trend in the data (Figure \ref{fig:SNR}b). The final two points (1150 $\mu$m and 1750 $\mu$m, respectively) are excluded from the linear regression as the extreme broadening removes any semblance of a gaussian resonance, producing an inaccurate fit (see Figures S7-S12).

A plot of the SNR versus the effective divergence ($\theta /2$) returns a volcano plot (Figure \ref{fig:SNR}c) where the SNR is found to be maximum (> 120) in the range of 0.4 to 0.8 milli-radians. The effective angular divergence is calculated by projecting a triangle from the point of divergence to the radius of the iris $r$, which are separated by a distance of 45 cm. $\frac\theta 2 = \tan ^{-1}\left(\frac{r}{45 \text{ cm}}\right)$. We propose that this region of the volcano plateau is the ideal entrance-slit configuration for the home-built XUV spectrometer. On the left edge of the volcano, higher resolution can be obtained with a steep loss in photon counts and, consequently, loss in SNR. The right edge, on the other hand, is significantly lop-sided and shows a simultaneous loss of both resolution and SNR. The asymmetry is captured in our ray-tracing simulations (Figure \ref{fig:RayTracing}d), noting the arctan relationship between iris size (in microns) and divergence (in milli-radians). Interestingly, the ionization edge remains identifiable up to the highest effective divergences sampled in the experiment. Under these conditions, the periodic structure in the high harmonic spectrum is no longer discernible, instead reflecting a broad continuum (Figure S5). A power law, non-resonant ionization background is also clearly seen for both the sample and substrate at >1.5 mrad angular divergence.

\section{Conclusion}
The results demonstrate that partially detuning the resolution provides increased photon flux to fill valleys between individual harmonic peaks and limits artificial modulations in the acquired absorption spectra.
Therefore, a motorized iris aperture in a diverging beam XUV spectrometer is used as an effective tool to provide tunability in the spectrometer resolution and SNR related to the overall photon flux. 
In metallic thin film samples with relatively broad XUV absorption features, a loss of spectral resolution from 600 meV to 1.8 eV does not alter the edge shape, providing an accurate measurement of the true absorption edge that is comparable to the CXRO database.
A volcano plot between the SNR and the projected divergence (inversely related to spectral resolution) exemplifies the tradeoff between these two critical spectroscopic descriptors for tabletop XANES measurements.
The on-demand control of photon flux and XUV divergence presents a handle for high SNR spectra, enabling faster data acquisition times at a lower exposure for improved throughput. The results pave the way for the rapid acquisition of single-shot XUV absorption spectra of thin film solid state samples on a tabletop source.

\section{Supplemental Information}

See Supplement 1 for supporting content.

The following files are available free of charge.
\begin{itemize}
  \item Figure S1-S2: SHADOW3 ray tracing simulations
  \item Figure S3-S5: Details of the iris aperture and harmonic spectra
  \item Figure S6-S12: Schematic harmonic spectrum and iron M2,3 spectrum analysis
\end{itemize}

\section{Acknowledgments}
This material is based upon work supported by the National Science Foundation under Grant No. 2440699.
{Initial results on the project} were supported by the donors of ACS Petroleum Research Fund under Doctoral New Investigator Grant 66711-DNI6. A.B. served as Principal Investigator on ACS PRF 66711-DNI6 that provided support for N.C.O and G.W.H.
The authors thank Josiah Wray, Austin Heidbreder, and Conrad Powell for their early contributions to beamline design and implementation (J.W.) and data collection (A.H. and C.P.).

\section{Disclosures}
The authors declare no potential conflicts of interest.

\printbibliography
\end{document}